\pdfoutput=1

\documentclass[11pt,a4paper]{article} 
\usepackage{jcappub} 

\usepackage{amsmath}
\usepackage{amsfonts}
\usepackage{graphicx}
\usepackage{bm}

\usepackage{varioref}
\usepackage{amssymb}
\usepackage{comment}

\title{Multi-field inflation with random potentials: field dimension, feature scale and non-Gaussianity}
\author{Jonathan Frazer}
\author{and Andrew R. Liddle} 
\affiliation{Astronomy Centre, University of Sussex, Brighton BN1 9QH,
United Kingdom}

\emailAdd{J.Frazer@sussex.ac.uk}
\emailAdd{A.Liddle@sussex.ac.uk}

\abstract{
We explore the super-horizon evolution of the two-point and three-point correlation functions of the primordial density perturbation in randomly-generated multi-field potentials. We use the Transport method to evolve perturbations and give full evolutionary histories for observables. Identifying the separate universe assumption as being analogous to a geometrical description of light rays, we give an expression for the width of the bundle, thereby allowing us to monitor evolution towards the adiabatic limit, as well as providing a useful means of understanding the behaviour in $f_{\rm{NL}}$. Finally, viewing our random potential as a toy model of inflation in the string landscape, we build distributions for observables by evolving trajectories for a large number of realisations of the potential and comment on the prospects for testing such models. We find the distributions for observables to be insensitive to the number of fields over the range 2 to 6, but that these distributions are highly sensitive to the scale of features in the potential. Most sensitive to the scale of features is the spectral index, with more than an order of magnitude increase in the dispersion of predictions over the range of feature scales investigated. Least sensitive was the non-Gaussianity parameter $f_{\rm{NL}}$, which was consistently small; we found no examples of realisations whose non-Gaussianity is capable of being observed by any planned experiment. 
}

\begin{document}

\maketitle
\flushbottom


\section{Why random potentials?}

\subsection{First reason}
With the prospect of improved data from the Planck mission fast approaching, there has been a lot of interest in finding inflationary models exhibiting specific observable footprints. Large non-Gaussianity signals peaking at various shapes is one example, another being features in the primordial power spectrum. While this is a crucial step towards understanding what observables are specific to a particular model, often the set-up can be somewhat contrived and to gain understanding as to whether such behaviour is a general feature of the model, one may need to invoke Monte Carlo techniques. In such a situation it may be helpful to employ some element of `randomness' at the level of the construction of the model. 

For example, a popular model of inflation coming from string theory is Dirac--Born--Infeld inflation. The DBI Lagrangian is
\begin{equation}
\mathcal{L}=-T(\phi)\sqrt{1-\frac{2 X}{T(\phi)}}+T(\phi)-V(\phi),
\end{equation}
where $X = -\frac{1}{2}g_{\mu\nu}\partial^{\mu}\phi\partial^{\nu}\phi$ and $T(\phi)$ is the brane tension. If the D3-brane velocity approaches its limiting speed
\begin{equation}
1-\frac{2 X}{T(\phi)}\rightarrow0,
\end{equation}
then a period of inflation can occur. This model has interesting observational consequences as the sound speed can become small and hence, since $f^{\rm{(eq)}}_{\rm{NL}}\sim 1/c_{\rm s}^2$, lead to large equilateral $f_{\rm{NL}}^{\rm{(eq)}}$. However the above Lagrangian also admits inflation by other means. In Ref.~\cite{Agarwal} a rather sophisticated model of brane inflation was investigated, where to simulate the effect of the bulk in different compactifications, random coefficients were used. In this set-up, conditions for DBI inflation were never encountered; instead inflection-point slow-roll inflation was vastly more common. We therefore see that while the DBI effect certainly gives an interesting observational footprint, there is no reason to believe this is a generic feature of brane models of inflation.\footnote{As pointed out in Ref.~\cite{Agarwal}, this result is not conclusive since, rather importantly, their investigation did not go all the way to the tip of the throat. Nevertheless we feel this example illustrates the point in hand.}

\subsection{Second reason}

On a more ambitious note, string theory seems to predict the existence of a landscape \mbox{\cite{HW,Susskind},} where, in the low-energy approximation, different regions may be characterised by the values of a large number of scalar fields. The consequence of this is that we have some very complicated potential $V(\phi_{1},...,\phi_{d})$ with a large number of minima each corresponding to a different metastable vacuum energy. This implies that instead of trying to predict the values of observables, we should be trying to predict probability distributions for them. Indeed, as we will now discuss, this is the case not just for string theory but for any model with multiple light fields. 

Most work on the consequences of a landscape has focussed on the measure problem (see Ref.~\cite{BenF} for a recent overview) but if the observational consequences of such a model are ever to be understood, then there are other challenges to contend with.  In order for a landscape model (any model where the scalar potential has more than one minimum, or for the purposes of this discussion, even just one minimum but multiple fields) to be predictive, three questions need to be addressed:
\begin{enumerate}
\item \emph{What are the statistical properties of the landscape}
\item \emph{What are the selection effects from cosmological dynamics}
\item \emph{What are the anthropic selection effects}
\end{enumerate}

The measure problem relates to the question of handling the numerous infinities which turn up. Taking the example of slow-roll inflation,\footnote{Most discussion in this area focuses on the scenario of inflation coming from tunnelling between metastable vacua but as we discuss here, the problem is much more general than that, affecting even the most pedestrian of inflationary set-ups.} any model of multi-field inflation suffers from an uncountably infinite set of choices for initial conditions. In general one needs to assume that, one way or another, at some point our region of spacetime experienced field values displaced from our local minimum. This corresponds to a single realisation of initial conditions (plus quantum scatter) that gave an anthropically suitable inflationary trajectory, which subsequently found its way to our local minimum. In order to make predictions we need to ask what proportion of the whole universe finds itself in this situation, i.e.\ what proportion of an infinite space finds itself in one of an infinite set of initial conditions. The ratio is ill-defined without a measure.

However overcoming this formidable task is not the end of it. Even with a solution to this measure problem we are still left with a considerable challenge. A solution to this issue is likely not to give us a specific set of initial conditions for a given model but a probability distribution for them. If all we can hope for is a statistical description of initial conditions, then in turn we only have a statistical description of inflationary trajectories and so, rather than calculating single values for observables, we should be calculating their distributions! The shape of these distributions will in part be determined by the model. This last point can, at least in some respects, be studied in its own right without a detailed knowledge of the string landscape or the measure problem. In this paper we take inspiration from the string landscape and study characteristics of these distributions in the context of a potential with multiple fields, containing a large number of vacua.

An early study of the possible consequences of this landscape picture for slow-roll inflation was carried out by Tegmark \cite{TegmarkInf}, who generated a large number of random one-dimensional potentials and explored the inflationary outcomes. In Ref.~\cite{me} we extended this to two fields to investigate the effect of entropy modes on super-horizon evolution. As already mentioned, in Ref.~\cite{Agarwal} a similar analysis was done for a six-field brane inflation model with random terms arising in the contribution coming from the bulk, where although the super-horizon effects were not analysed, both reassuringly and rather excitingly, qualitatively similar emergent behaviour was identified to that found in Ref.~\cite{me}. In this paper we further extend our work in Ref.~\cite{me} to a larger number of fields and a broader range of potentials, as well as obtaining results for the non-Gaussianity $f_{\rm{NL}}$. The aim is to gain insight into the origin and limits of emergent behaviour.

\section{Architecture}

\subsection{Not the most general Fourier series}

We construct our potential following an approach similar to Refs.~\cite{TegmarkInf,me}. We use a random function of the form
\begin{equation}
V(\phi)  = m^4_{{\rm v}}\sum_{1<k_{i}<k_{\rm max}} \left[a_{\bf{k}} \cos{\left(\frac{\bf{k.\phi}}{m_{{\rm h}}}\right)}+b_{\bf{k}} \sin{\left(\frac{\bf{k.\phi}}{m_{{\rm h}}}\right)}\right]
\end{equation}
where $\rm{\phi}$ is the vector $\phi_{i}$ with $i$ running from 1 to $d$, as is the vector $\bf{k}$ and $m_{{\rm v}}$ and $m_{{\rm h}}$ are the vertical and horizontal masses respectively. The summation in $k_{i}$ means $d$ summations take place where in each case the summation runs from $k_{i}=1$ to $k_{i}=k_{\rm{max}}$. The amplitudes $a_{\bf{k}}\equiv a_{k_{1},...,k_{d}}$ are independent Gaussian random variables with zero mean and standard deviation
\begin{equation}\label{eq:coef}
\sigma = e^{-{\bf k.k}/{k_{{\rm max}}}\,d },
\end{equation}
Due to computational limitations we cannot make $k_{\rm max}$ sufficiently large to see the effect of the central limit theorem come into play. So rather than having a variance of order unity, we control the variance in the above manner. While this is a helpful thing to do computationally, one needs to bear this in mind when considering principles of effective field theory. We will return to this discussion shortly.

The potentials we simulate are periodic with periodicity scale $2\pi m_{\rm{{\rm h}}}$, and we can only expect reasonable results if the field trajectory spans a distance in field space less than the periodicity of the random function. This turns out to always be the case.

\begin{figure}[t]
\centering
\includegraphics[width=15cm]{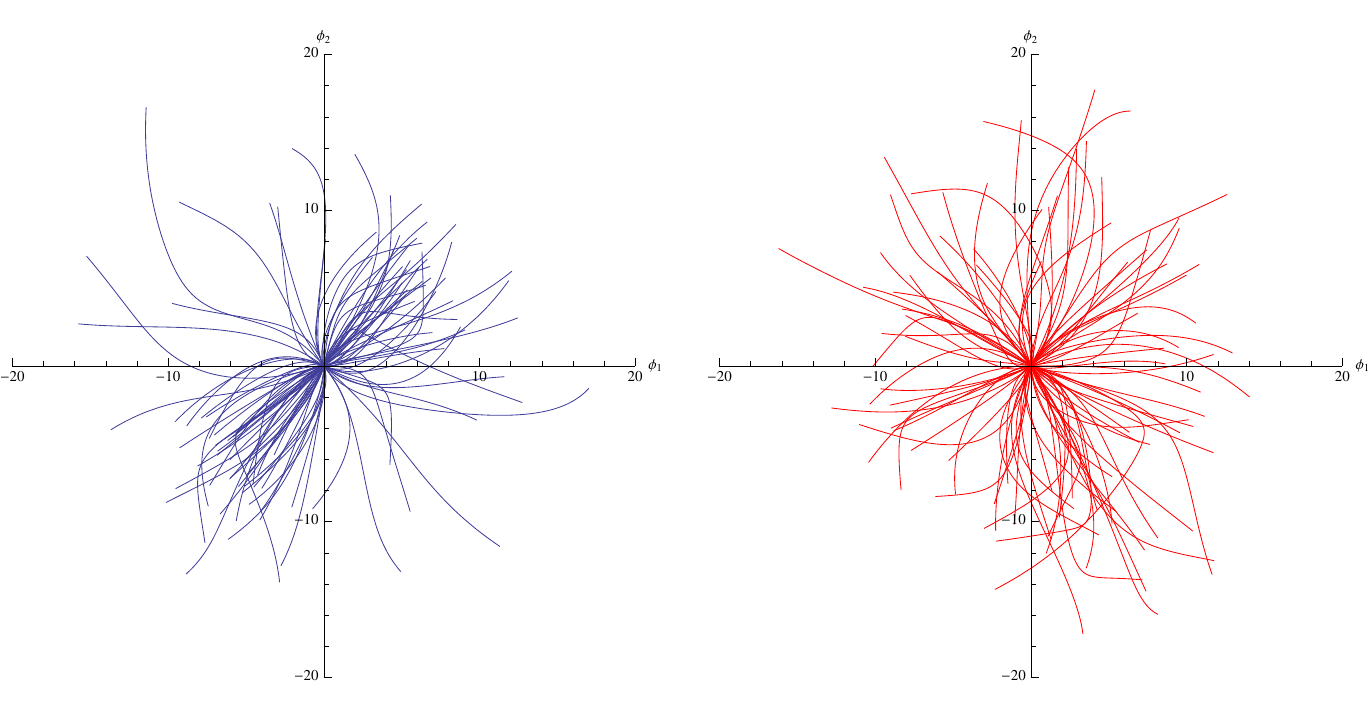}
\caption{Example trajectories for our potential (blue) and Fourier series (red) for two-field model with $k_{\rm max}=3$.}
\label{fig:trajectories}
\end{figure}

Note that summing the potential this way means we are not using the most general Fourier series. As shown in Fig.~\ref{fig:trajectories}, by restricting the summations over each $k_{i}$ to non-negative values we sacrifice statistical $d$-spherical symmetry but in doing so we are able to build an observationally indistinguishable potential out of a fraction of the number of terms (see Table \ref{t:terms} for the number of terms in the series for various $d$ and $k_{\rm max}$ values). 

\subsection{Experiment-specific considerations}

As discussed, any model of inflation where the potential has multiple minima predicts a probability distribution for the cosmological parameters. We wish to compute this distribution for various potentials of the above form. To do this we perform the same experiment as that performed in Refs.~\cite{TegmarkInf, me}: 
\begin{enumerate}
\item Generate a random potential $V(\mathbf{\phi})$ and start at $\mathbf{\phi}=(0,0)$.
\item If $V(0,0)<0$ then reject model, otherwise evolve to find the field trajectory.
\item If model gets stuck in eternal inflation, reject.
\item Once the model stops inflating, if the number of e-folds of inflation $N<60$ we reject as insufficient inflation occurred, otherwise calculate observables. 
\item Repeat steps 1-4 many times to obtain a statistical sample.
\item (Change some assumptions and do it all again.)
\end{enumerate}
Note that due to our potential being statistically invariant under translation, generating multiple realisations of the potential and starting at the origin is equivalent to taking a single realisation and scanning over initial conditions.

\begin{table}[t]
\centering
\begin{tabular}{|c|c|c|c|c|c|c|c|c|c|c|c|c|}\hline
\multicolumn{1}{|c|}{ } & \multicolumn{6}{|c|}{Our Potential} & \multicolumn{6}{|c|}{Fourier Series} \\ \hline
 Fields   $d$      & 2  & 2  & 4  & 4  & 6  & 6 & 2  & 2  & 4  & 4  & 6  & 6   \\ \hline
 $k_{\rm{max}}$ & 3  & 5 &  3 & 5  &  3 & 5 & 3  & 5 &  3 & 5  &  3 & 5 \\ \hline
 $\#$ of terms         &18   & 50 & 162  & 1250 & 1458 &31250   & 98 & 242  & 4802 &29282   & 235298 & many   \\ \hline 
    \end{tabular}
\caption{Summary of how the number of terms in the potential changes  with the truncation $k_{\rm max}$ and number of fields $d$ for Fourier series potential and our reduced version.}
\label{t:terms}
\end{table}

In Ref.~\cite{me}, taking the final minimum as the ultimate vacuum energy, to give an approximately anthropically suitable solution \cite{Weinberg} we had an additional cut stipulating the final vacuum energy must be positive to avoid subsequent collapse. This, in conjunction with the rejection of eternally inflating vacua, was found to be an extremely severe cut, in some cases reducing the proportion of otherwise viable solutions from 0.06 to more like $2\times 10^{-5}$. In this paper we abandon this cut,  to enable us to explore more featured potentials which would not otherwise be computationally accessible. We found this to be of little consequence for observables. An explanation for this is that the two models may differ only in the nature of the post-inflationary evolution of the trajectory, which has no effect on the evolution of observable quantities. 

The other consideration regarding the experimental set-up is at what value to set the vertical and horizontal mass scales $m_{{\rm v}}$ and $m_{{\rm h}}$. The vertical mass has little dynamical impact and only affects the amplitude of the observed power spectrum by a factor and not other observables. For this reason, rather than fixing $m_{{\rm v}}$ we adjust it on a case-by-case basis such that the amplitude at horizon exit is $P^{*}_{\zeta}=2\times 10^{-5}$. 

\begin{figure}[t]
\centering
\includegraphics[width=14.5cm]{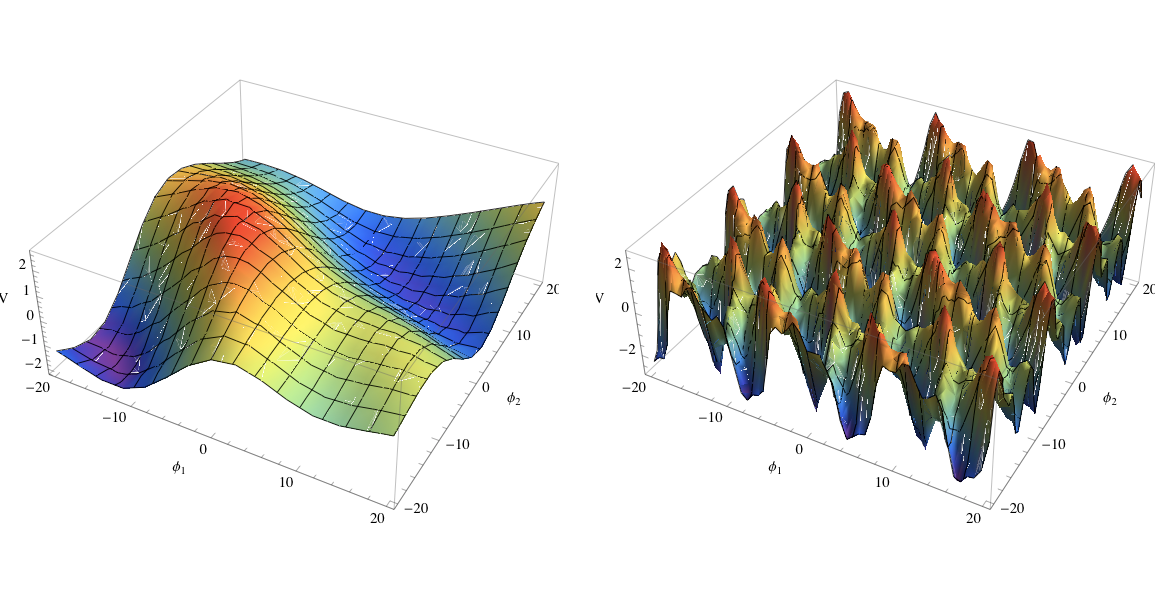}
\caption{Example of two-field potentials with $m_{{\rm h}}=15.8\,M_{\rm{Pl}}$ and $m_{{\rm h}}= 2.0\,M_{\rm{Pl}}$ respectively.}
\label{fig:mhsmaller}
\end{figure}

The horizontal mass $m_{{\rm h}}$ is more interesting. As previously discussed, motivated by the aim of minimising the number of terms in the potential for a given dynamical behaviour, the random coefficients are chosen in such a way as to make the potential essentially insensitive to truncation. This set-up means that $m_{{\rm h}}$ is our key parameter in adjusting how featured the potential is. As the examples in Fig.~\ref{fig:mhsmaller} show, adjusting the scale of features affects the length scale $\Delta\phi$ of the inflationary distance in field space. We will discuss motivation from theory for this length scale next, but it is important we understand its implications for predictability and thus we shall be showing results for a range of $m_{{\rm h}}$ values.

\subsection{Basic considerations from effective field theory}\label{sec:fieldth}

Generally one expects that the inflationary potential can be well described by an effective theory containing non-renormalizable contributions coming from integrating out massive fields.\footnote{Discussion along these lines can be found in Ref.~\cite{danbmca}  and of course Ref.~\cite{Lythbound} but we would particularly like to thank Liam McAllister and Sam Rogerson for very helpful clarifications and additional comments on this matter.} For instance, for a single-field model one can write
\begin{equation}\label{eq:polyV}
V(\phi)=V_{0}+\frac{1}{2}m^{2}\phi^{2}+M\phi^{3}+\frac{1}{4}\lambda\phi^{4}+\sum_{d=5}^{\infty}\lambda_{d}M_{\rm{Pl}}^{4}\left(\frac{\phi}{M_{\rm{Pl}}}\right)^d,
\end{equation}
where the terms in the summation are non-renormalizable. One expects the masses in the summation to be at or even well below the Planck mass as, in analogy to the argument from $W$--$W$ scattering for a Higgs around 1 TeV, there needs to be something to unitarize graviton--graviton scattering. There is no good reason to assume the inflaton does not couple to these extra degrees of freedom. To do so is to make a strong assumption about quantum gravity which is hard to justify, and thus we expect $\lambda_{d}\sim 1$. If we categorise inflation models as large field, $|\Delta \phi|\gg M_{\rm{pl}}$, medium field $|\Delta \phi|\sim M_{\rm{pl}}$ and small field  $|\Delta \phi|\ll M_{\rm{pl}}$, then this sort of reasoning indicates small and perhaps medium field models should be considered more realistic as terms in the summation are suppressed, while to have a large-field model, one needs to justify additional symmetries to protect the flatness of the potential against the otherwise increasingly large series contributions. Crudely speaking we can think of our choice of $m_{{\rm h}}$ as corresponding to a decision on what energy scales we are integrating out.

Finally, we would like to consider the number of fields to be included in the model. Historically a lot of focus has been given to single-field models simply because they are the most basic inflationary set-up, but this is not what is best motivated from the field theory perspective. As already mentioned, a single-field model occurs when one degree of freedom is much lighter than all the others. This means one can integrate out the other degrees of freedom provided they are sufficiently massive, but there is no good reason to believe this is necessarily the case. For example, in string theory the contributing massive fields include stabilised moduli. Work on flux compactifications is still very much in development but typically masses correspond to around the Hubble scale. This strongly motivates models with tens if not hundreds of active fields \cite{CBandL}. 

It therefore seems quite reasonable to model the final inflationary phase in a landscape as a truncated $d$-field Fourier series with random coefficients, provided we are dealing with small- to medium-field models. However, for computational reasons we are forced to work with something less realistic. Ideally we would work with more fields, and push to smaller field excursions than we will be working with. 
For the purposes of our investigation we will at times be working with inflationary trajectories that not particularly well motivated as genuine models of inflation, yet we still find them to be quite informative when it comes to understanding inflationary dynamics.

\section{The Transport equations}

In this paper, we improve on our previous work \cite{me} by calculating the perturbations using the Transport method of Mulryne et al.~\cite{MT1,MT2}. This gives improved computational efficiency for the power spectrum, while still including all isocurvature effects, and additionally allows us to compute the non-Gaussianity parameter $f_{\rm NL}$. We compared results from this method with the geometric approach (see Ref.~\cite{geometric} for early work in this area; see Ref.~\cite{geomrecent} for some more recent work) used in our previous paper \cite{me}, and they were found to agree for all models tested, as well as giving the same distributions of observables when tested on our landscape model. 

With regard to calculating $f_{\rm{NL}}$, compared to other methods (for instance Ref.~\cite{deltaN,WandsC}), the Transport approach has the benefit of being computationally more efficient, as well providing a new means of understanding contributions to $f_{\rm{NL}}$ by having explicit source terms. Equivalent to all other methods in the literature (including cosmological perturbation theory), it is simply an implementation of the separate universe assumption, but instead of evolving many perturbed trajectories, as is done in the popular $\delta N$ approach \cite{deltaN, WandsC,deltaNearly}, one evolves probability distributions.

What follows is largely a summary of the work done in Refs.~\cite{MT1,MT2}. We focus on explicitly showing how the Transport formalism is implemented for a general $d$-field model of inflation and refer the reader to Refs.~\cite{MT1,MT2} for the details.

\subsection{Moments of $\zeta$}\label{sec:moments}

In calculating the statistical properties of the curvature perturbation we invoke the separate universe assumption and consider a collection of space-time volumes whose mutual scatter will ultimately determine the microwave background anisotropy on a given scale. Each space-time volume follows a slightly different trajectory in field space, whose position at a given time we label $\phi^*$, the scatter of which is determined by the vacuum fluctuations at horizon exit. Here and in what follows the superscript ``$*$" indicates that the quantity is evaluated on a spatially-flat hypersurface. If we know the distribution $P(\phi^*)$ then, among other things, we can study the statistical properties of the deviation of these trajectories from their expectation value $\Phi$, $\delta \phi_{i}^*=\phi_{i}^*-\Phi_{i}^* $, where $i$ indexes the components of the trajectory $\phi$, namely the species of light scalar fields. The two-point correlations among the $\delta\phi_{i}$ are expressed by the covariance matrix $\Sigma (t)$, where
\begin{equation}\label{eq:sigma}
\Sigma_{ij} \equiv \langle\delta\phi_{i}\delta\phi_{j}\rangle 
\end{equation}
and the third moment is given by 
\begin{equation}\label{eq:alpha}
\alpha_{ijk} \equiv \langle\delta\phi_{i}\delta\phi_{j}\delta\phi_{k}\rangle 
\end{equation}
The covariance matrix, third moment and centroid $\Phi$ are all functions of time, but in our notation we will be suppressing the explicit time dependence. 

A consequence of the separate universe assumption \cite{deltaNearly} is that the curvature perturbation $\zeta$ evaluated at some time $t=t_{c}$ is equivalent on large scales to the perturbation of the number of e-foldings $N(t_{c},t_{*},x)$ from an initial flat hypersurface at $t=t_*$, to a final uniform-density hypersurface at $t=t_c$,
\begin{equation}
\zeta(t_c,x) \simeq \delta N(t_c,t_*,x)\equiv N(t_c,t_*,x)-N(t_c,t_*)
\end{equation}
where
\begin{equation}
N(t_c,t_*) \equiv \int_{*}^{c}H dt.
\end{equation}
Expanding $\delta N$ in terms of the initial field perturbations to second order, one obtains
\begin{equation}\label{eq:zeta}
\zeta(t_c,x)=\delta N(t_c,t_*,x)=N_{,i}\delta\phi_i^* +\frac{1}{2}N_{,ij}(\delta\phi_{i}\delta\phi_{j}-\langle\delta\phi_{i}\delta\phi_{j}\rangle),
\end{equation}
where repeated indices should be summed over, and $N_{,i}$, $N_{,ij}$ represent first and second derivatives of the number of e-folds with respect to the fields $\phi_{i}^*$. We remind the reader that it is necessary to subtract the correlation function in the second term. This is because one can interpret the covariance matrix as the contribution from disconnected diagrams which gives the vacuum energy. In Fourier space one only considers connected diagrams from the outset and thus the subtraction is already implicitly taken care of.

Combining Eq.~\eqref{eq:sigma} and Eq.~\eqref{eq:alpha} with Eq.~\eqref{eq:zeta} we get expressions for the two- and three-point functions in terms of the moments of $P(\phi^{*})$. The two-point function is
\begin{equation}\label{eq:2p}
\langle\zeta\zeta\rangle=N_{,i}N_{,j}\Sigma_{ij}.
\end{equation}
It is useful to decompose the three-point function as
\begin{equation}
\langle\zeta\zeta\zeta\rangle=\langle\zeta\zeta\zeta\rangle_{1}+\langle\zeta\zeta\zeta\rangle_{2},
\end{equation} 
where 
\begin{equation}\label{eq:3p1}
\langle\zeta\zeta\zeta\rangle_{1}=N_{,i}N_{,j}N_{,k}\alpha_{ijk},
\end{equation} 
and
\begin{equation}\label{eq:3p2}
\langle\zeta\zeta\zeta\rangle_{2}=\frac{3}{2}N_{,i}N_{,j}N_{,km}\left[\Sigma_{ik}\Sigma_{jm}+\Sigma_{im}\Sigma_{jk}\right].
\end{equation} 
Eq.~\eqref{eq:3p1} is the intrinsic non-linearity among the fields, while Eq.~\eqref{eq:3p2} encodes the non-Gaussianity resulting from the gauge transformation to $\zeta$; as one evolves from one flat hypersurface to another, turns in the trajectory will contribute to the non-Gaussianity. This, as well as any non-Gaussianity present at horizon exit, is what is encapsulated in Eq.~\eqref{eq:3p1}. However, this super-horizon evolution also causes the hypersurface of constant density to change and so the gauge transformation from the flat hypersurface to the coinciding surface of constant density also contributes to the non-Gaussianity and this contribution is taken into account in Eq.~\eqref{eq:3p2}.

\subsection{Derivatives of $N$}

From Eq.~\eqref{eq:3p1} and Eq.~\eqref{eq:3p2}, it is clear that in order to calculate moments of the power spectrum we need a method for calculating derivatives of $N$. In general when using the $\delta N$ technique it is difficult or impossible to find an analytic expression for the derivatives of $N$.  It is therefore necessary to run the background field equations many times from perturbatively different initial conditions, stopping at some value for $H$ which is the same for all the runs.  One then calculates the derivatives of $N$ with respect to the initial conditions. In using the Transport equations, however, this process is replaced by solving a set of coupled ordinary differential equations. Instead of taking the surfaces ``$*$'' and ``$c$" to be at horizon crossing and time of evaluation\footnote{Time of evaluation is often taken to be the end of inflation but with regard to calculating observables, any time after isocurvature modes have decayed away will give the same result. A problem arises when isocurvature modes are still present at the end of inflation. In this case the power spectrum will continue to evolve and without a model of reheating this renders the model non-predictive. We will return to this in greater detail later on.}  respectively, instead the surfaces are taken to be infinitesimally separated and the transport equations evolve the field values at horizon crossing forward to the time of evaluation. The upshot of this is two-fold. As we will see, the use of ordinary differential equations to evolve the moments of the field perturbations allows us to see the source of super-horizon evolution and hence the various contributions to $f_{\rm NL}$. The second and more immediate benefit to our current discussion is that we can find a general expression for the derivatives of $N$. To leading order in slow-roll, for a given species ``$i$", the number of e-folds $N$ between the flat hypersurface and a comoving hypersurface is given by
\begin{equation}
N(t_c,t_*) \equiv -\int_{\phi^*}^{\phi^c}\frac{V}{V_{,i}} d\phi_{i}\,, \qquad \mbox{no sum on $i$} \,.
\end{equation}
and so if the two surfaces are infinitesimally separated, then we can write 
\begin{equation}
dN=\left[\left(\frac{V}{V_{,i}}\right)^{*}-\left(\frac{V}{V_{,j}}\right)^{c}\frac{\partial \phi^{c}_{j}}{\partial \phi^{*}_{i}}\right] d\phi_{i}^{*}.
\end{equation}
To handle $\partial \phi_{j}^{c}/\partial\phi_{i}^{*}$ the method used in Refs.~\cite{WandsC, BandEC} for sum-separable potentials is also now applicable and we introduce the quantity  
\begin{equation}
C_{i} \equiv -\int\frac{d\phi_{i}}{V_{,i}} +\int\frac{d\phi_{i+1}}{V_{,i+1}} .
\end{equation}
This enables us to write
\begin{equation}
d\phi_{i}^{c}=\frac{\partial \phi^{c}_{i}}{\partial C_{j}}\frac{\partial C_{j}}{\partial \phi_{k}^{*}} d\phi_k^{*}
\end{equation}
which after some algebra gives the expression
\begin{equation}
\frac{\partial\phi_{i}^{c}}{\partial\phi_{j}^{*}}=-\left(\frac{V}{V_{,j}}\right)^{*}\left(\frac{V_{,i}}{V}\right)^{c}\left(\frac{V_{,j}^{c}}{V_{,k}^{c}V_{,k}^{c}}-\delta_{ij}\right)
\end{equation}
Hence we find
\begin{equation}
N_{,i}=\left(\frac{V}{V_{,i}}\right)^{*}\left(\frac{V_{,i}^{2}}{V_{,k}V_{,k}}\right)^{c},  \qquad \mbox{no sum on $i$}
\end{equation}
and
\begin{equation}\label{eq:Nij}
N_{,ij}=\frac{V_{,i}V_{,j}}{V_{,k}V_{,k}}+\frac{V V_{,ij}}{V_{,k}V_{,k}}-\frac{2V V_{,i k}V_{,k}V_{,j}}{(V_{,k}V_{,k})^2}-\frac{2V V_{,j k}V_{,k}V_{,i}}{(V_{,k}V_{,k})^2}+\frac{2V V_{,i}V_{,k}V_{,k l}V_{,l}V_{,j}}{(V_{,k}V_{,k})^3},
\end{equation}
where in Eq.~\eqref{eq:Nij} the limit $c\rightarrow*$ has been taken.

\subsection{Transporting the moments}

Finally, we need a method for evolving the moments of the scalar perturbations. The probability distribution $P(\phi^*)$ is conserved and so, as described by the standard continuity equation, the rate of change of $P$ is given by the divergence of the current,
\begin{equation}\label{eq:continuity}
\frac{\partial P}{\partial N}+\frac{\partial(u_{i}P)}{\partial \phi_{i}}=0,
\end{equation}
where $u_{i}\equiv \phi'_{i}$ is the field velocity. The key achievement of Ref.~\cite{MT1} was to develop a method for extracting the evolution equations of the moments of $P$ from the continuity equation. In Ref.~\cite{MT2} an alternative method was introduced, generalising to any number of fields on arbitrary slicing. We do not go into the techniques here; instead we just quote the resulting evolution equations for the centroid, variance and skew which collectively we refer to as the Transport equations, 
\begin{equation}\label{eq:centroid}
\Phi'_{i}=\phi'_{i}+\frac{1}{2}u_{i,mn}\Sigma_{mn}+...\ ,
\end{equation}
\begin{equation}\label{eq:variance}
\Sigma'_{ij}=u_{i,m}\Sigma_{mj}+u_{j,m}\Sigma_{mi}+\frac{1}{2}u_{i,mn}\alpha_{jmn}+\frac{1}{2}u_{j,mn}\alpha_{imn}+...\ ,
\end{equation}
\begin{equation}\label{eq:skew}
\alpha'_{ijk}=u_{i,m}\alpha_{mjk}+u_{i,mn}\Sigma_{jm}\Sigma_{kn}+({\rm cyclic }\   i\rightarrow j\rightarrow k)+ \cdots\ .
\end{equation}
The equation for the centroid Eq.~\eqref{eq:centroid} says that the mean field value evolves as the velocity of the fields but can be affected by evolution of the wings of the distribution. The evolution equations for the variance and skew, as one might guess from the continuity equation Eq.~\eqref{eq:continuity}, give evolution as the divergence of the field velocity but now also with source terms coming from the other moments.

\subsection{Cross-sections of the bundle}

As will be discussed in more detail in due course, an important consideration in our analysis will be whether or not evolution of observables is still taking place at the time of evaluation. Evolution stops when the trajectory becomes effectively single field \cite{GBW}. This is to say the trajectory has reduced to a caustic \cite{JDD}, so for this reason we would like a description for the evolution of the cross-section of the perturbed trajectories. Such a description has recently been developed in Ref.~\cite{ray}, to which we refer the reader for more detailed discussion. For simplicity we only describe the broad concept here and quote results that will be needed in future discussion.

Cross-sections within the bundle are focused, sheared and rotated by the flow. These distortions can be characterised by the evolution of connecting vectors describing the displacement between nearby trajectories in the bundle. If $\delta x_{i}$ is an infinitesimal connecting vector, then assuming $u_{i,j}$ is sufficiently smooth, $\delta x_{i}$ is transported as
\begin{equation}
\dot{\delta x}_{i}=\delta x_{j}\frac{\partial u_{i}}{\partial \phi_{j}}
\end{equation}
It follows that changes in the cross-section of the bundle can be determined in terms of the expansion tensor $u_{i,j}$. We can decompose this in terms of a dilation $\theta={\rm tr}\,u_{i,j}$, a traceless symmetric shear $\sigma_{ij}$, and a traceless antisymmetric twist $\omega_{ij}$,
\begin{equation}
u_{i,j}\equiv\frac{\theta}{d}\delta_{ij}+\sigma_{ij}+\omega_{ij}
\end{equation}

Dilation describes a rigid rescaling of $\delta x_{i}$ by $1+\theta$, representing a global tendency of the trajectories to focus or defocus. The shear encapsulates the tendency for some trajectories to flow faster than others while conserving the cross-sectional area of the bundle. The twist represents a rotation of the bundle with preserved volume, such as the tendency of trajectories to braid. The dilation, shear and twist act as sources for one another and so one expects a bundle will typically exhibit all of these behaviours at some point.

We refer the reader to Ref.~\cite{ray} for a more formal description of this formalism and its applications, but for the purposes of this paper all we need is the result that the focussing of the bundle is given by
\begin{equation}\label{eq:thetah}
\Theta(H,H_{0})=\exp \left[ \frac{1}{d}\int^{H}_{H_{0}}\theta(h)dh \right].
\end{equation}

\section{Findings}

Having set up our models and the machinery necessary to compute the observables, we now proceed to our results. The principal variables of interest to vary are the number of fields $d$ and the horizontal mass scale $m_{{\rm h}}$. Large values of the latter correspond to relatively smooth potentials, and small values to heavily featured potentials. We refer to individual realizations giving sufficient inflation as `verses'.

We discuss our results in the following sequence:
\begin{enumerate}
\item Dynamical properties of trajectories.
\item Perturbation evolution along individual trajectories.
\item Distribution of observables over ensembles of trajectories.
\end{enumerate}

\subsection{Trajectory dynamics}

\begin{figure}[t]
\centering
\includegraphics[width=15cm]{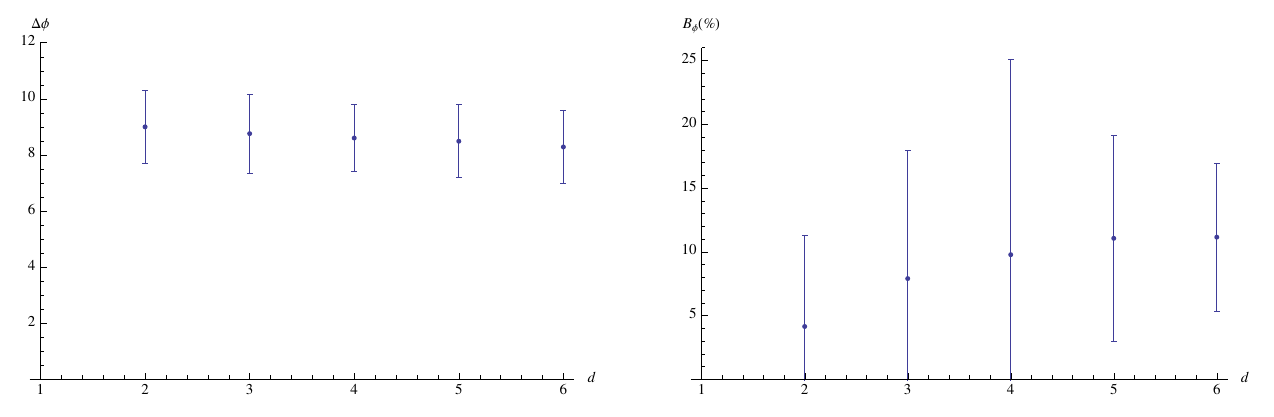}
\caption{How the mean $\Delta\phi$ and percentage increase in $\Delta\phi$ due to bending of the trajectory changes with the number of fields, for $m_{{\rm h}}=15\, M_{\rm{Pl}}$. The bars show how the standard deviation of these quantities changes over this range.}
\label{fig:dphid}
\end{figure}

\paragraph{Minimal $d$ dependence}Fig.~\ref{fig:dphid} summarises the qualitative behaviour found during our exploration of the properties of multi-field trajectories. We see that $\Delta\phi $, the length of the trajectory in field space, and $B_\phi$, the percentage increase in $\Delta\phi$ due to turns in the trajectory, defined as
\begin{equation}
B_\phi\equiv 100 \frac{\Delta\phi-\sqrt{\phi^{\rm end}\cdot\phi^{\rm end}-\phi^{*}\cdot\phi^{*}}}{\Delta\phi}
\end{equation}
show only a mild sensitivity to the number of fields. In fact, for all observables and inflationary parameters we looked at, the sensitivity to changing the number of fields was small over the range $d=2$ to $d=6$ compared to the spread of results. This was true even for the relatively predictive large-field case of $m_{{\rm h}}=15\,M_{\rm{Pl}}$. Least sensitive of all, we found no change whatsoever in the distribution of slow-roll parameters at horizon crossing with  $\epsilon^{*}=0.002\pm0.002$ and $\eta^{*}=0.0012\pm0.0004$. 

\paragraph{e-fold distributions} One instance where we did find sensitivity to $d$ was in the e-fold distributions, where we saw a decrease in the proportion of trajectories with more than $60$ e-folds from 0.08 for two fields down to 0.01 for six fields. Fig.~\ref{fig:dphid} hints at a tendency for trajectories to become shorter and more curved as the number of fields increases. So given that the trajectory will seek the route of steepest descent, it appears that for our model, increasing the number of fields increases the chance of the trajectory encountering a slope sufficiently steep to kill inflation.

In Refs.~\cite{Agarwal, Susskind} slightly different parameterisations of the inflationary region of the potential were used to show the probability of obtaining a given number of e-folds of inflation was 
\begin{equation}\label{eq:PN}
P(N)\propto\frac{1}{N^\alpha},
\end{equation}
where $\alpha=4$ and $\alpha=3$ were found respectively. Bearing Eq.~\eqref{eq:polyV} in mind, it seems reasonable that  Eq.~\eqref{eq:PN} might apply more generally and indeed we find it to be a good fit to our e-fold distributions, with $\alpha$ increasing as the potential becomes more featured from $\alpha=2$ through to $\alpha=5$ over the range of $m_{{\rm h}}$ we investigated. To illustrate the implication of this, consider the popular idea that inflation was preceded by a tunnelling event. Using the (perhaps somewhat arbitrary) values of Ref.~\cite{Susskind}, we place an anthropic lower bound on the number of e-folds at 59.5 coming from structure formation and an observational lower bound at $N = 62$ on the curvature from tunnelling. Then for $\alpha = 2$, the probability of the model achieving sufficient inflation to be in agreement with observation is roughly 92\%, while for $\alpha = 5$ it is more like 81\%. However, remember we are working in the range of $\Delta\phi$ which is not best motivated theoretically. While not accessible with the techniques used here, our results lead us to expect that for small-field models, $\alpha$ should be larger. This has the potential to cause tension with observation, as by $\alpha=18$ the chance of finding ourselves in the observed universe falls to $49\%$, i.e a typical observer would expect to see evidence of curvature.

\subsection{Perturbation evolution: $P_{\zeta\zeta}$, $f_{\rm{NL}}$ and the adiabatic limit}

Having seen that our model is insensitive to the number of light fields, for simplicity we only give results for two-field potentials in the remaining sections of this paper, focussing mainly on the dependence on the feature scale of the potential. But we would like to emphasise that the results hold more generally. 

\subsubsection{$P_{\zeta\zeta}$ and $\Theta$}

The key difference between single-field and multi-field inflation is that the latter admits evolution of the power spectrum on super-horizon scales. This means that in order to make a prediction from multi-field inflation one needs to know the full evolutionary history up until the model becomes effectively single field, i.e.\ the adiabatic limit is reached \cite{GBW, JDD}. Once this happens the power spectrum stops evolving and one can evaluate observable quantities at a subsequent time of one's pleasing. The problem is that there is no guarantee that such an adiabatic limit will be reached before the end of inflation, and if this is not the case making a prediction requires knowledge of reheating and so forth. 

\begin{figure}[t]
\centering
\includegraphics[width=15cm]{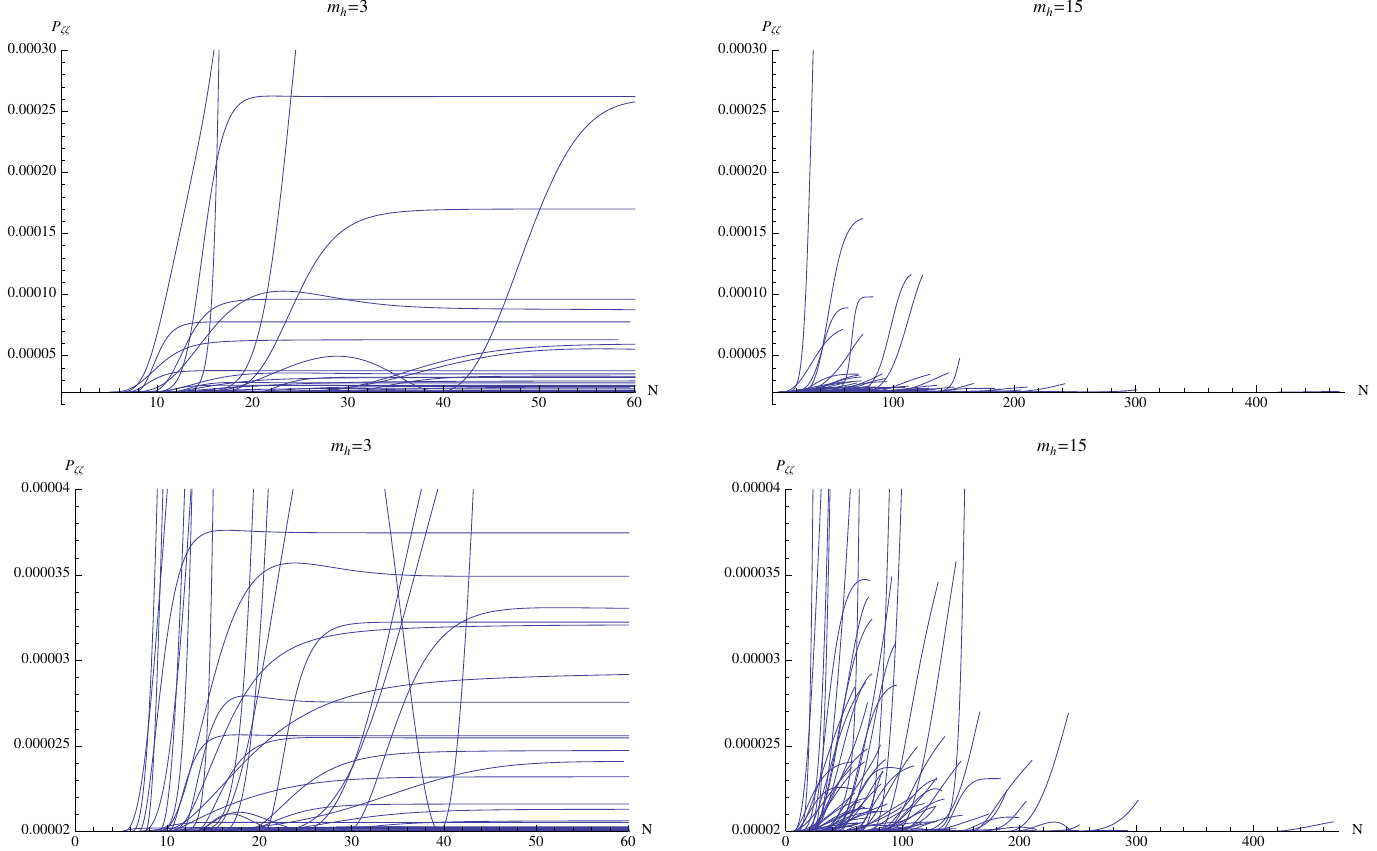}
\caption{Example plots showing the super-horizon evolution of the power spectrum for modes exiting 55 e-folds before the end of inflation. $m_{{\rm h}}=3 M_{\rm{Pl}}$ corresponds to a highly-featured potential, and in all such cases evolution stops long before the end of inflation. $m_{{\rm h}}=15 M_{\rm{Pl}}$ is a comparatively smooth landscape, and then a significant proportion of verses are still evolving at the end of inflation.}
\label{fig:pwr}
\end{figure}

\begin{figure}[t]
\centering
\includegraphics[width=15cm]{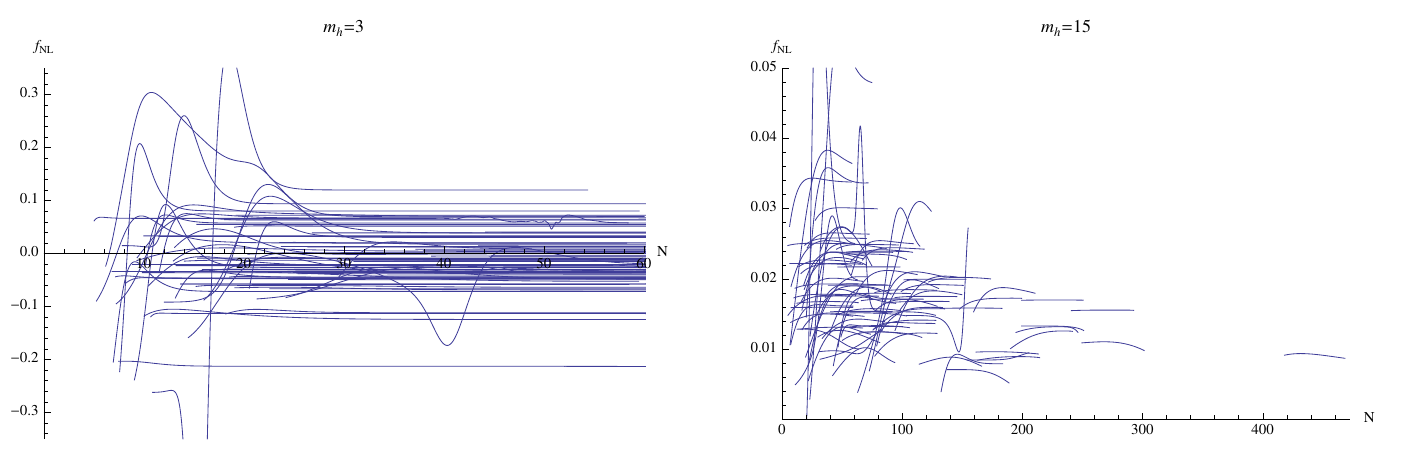}
\caption{Example plots showing the super-horizon evolution of $f_{\rm{NL}}$ for modes exiting 55 e-folds before the end of inflation. Again we see that for the more featured landscape evolution stops early on, while for the smoother example, evolution often continues to the end of inflation.}
\label{fig:fnlevol}
\end{figure}

\begin{figure}[t]
\centering
\includegraphics[width=10cm]{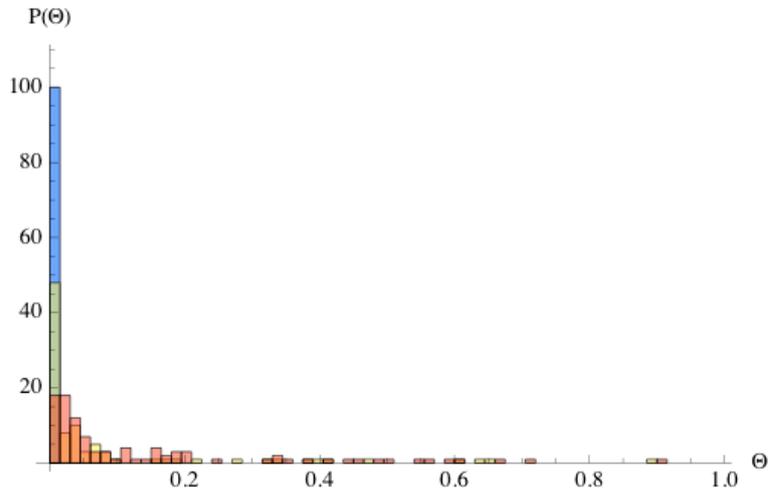}
\caption{$\Theta$ at the end of inflation for $m_{{\rm h}}=3\, M_{\rm{Pl}}$ (blue), $m_{{\rm h}}=9\, M_{\rm{Pl}}$ (yellow) and $m_{{\rm h}}=15\, M_{\rm{Pl}}$ (red). All trajectories essentially reach a caustic in the most featured example, less for $m_{{\rm h}}=9\, M_{\rm{Pl}}$ and least for the smoothest landscape $m_{{\rm h}}=15\, M_{\rm{Pl}}$. }
\label{fig:Theta}
\end{figure}

As can be seen in Figs.~\ref{fig:pwr}, \ref{fig:fnlevol} and \ref{fig:Theta}, the ability to reach the adiabatic limit, where the perturbation on a given scale becomes constant, is strongly dependent on how featured the landscape is.  For the more featured landscapes such as $m_{{\rm h}}=3\, M_{\rm{Pl}}$, we found the adiabatic limit was reached in all cases (even the trajectories disappearing off the top of the plots), while for the smoother landscapes like $m_{{\rm h}}=15 M_{\rm{Pl}}$ the proportion of trajectories achieving this clearly drops significantly. 

There is a very intuitive reason for why this should be the case. Rewriting Eq.~\eqref{eq:thetah} in terms of e-folds $N$ we find
\begin{equation}\label{eq:thetaN}
\Theta(N,N_{*})=\exp \left[\frac{1}{d}\int^{N}_{N_{*}}(3\epsilon-2\bar{\eta}+{\rm tr}\,M_{ij}) dN \right]
\end{equation}
where $M_{ij}$ is the Hessian of $\ln V$ and $\bar{\eta}$ is the generalised slow-roll parameter
\begin{equation}
\bar{\eta}\equiv\frac{V_{,i}V_{,j}V_{,ij}}{VV_{,k}V_{,k}}
\end{equation}
We therefore see that in a valley, strong focussing will occur, while on a ridge or a hilltop the bundle will dilate. With this picture it is quite easy to see why we should expect evolution as seen in Fig.~\ref{fig:pwr}. The treacherous landscape of $m_{{\rm h}}=3\, M_{\rm{Pl}}$ typically gives exactly the conditions required for a very strong focusing, while in contrast the comparatively mild, undulating meadows of $m_{{\rm h}}=15\, M_{\rm{Pl}}$ give very little incentive for trajectories to focus to a caustic.

\subsubsection{$f_{\rm{NL}}$ is always small}

Much of this kind of discussion carries over to understanding the results of Fig.~\ref{fig:fnlevol}. First and foremost it should be noted there was not a single example of a trajectory that gave sufficient non-Gaussianity to be detected by any future planned experiment. Methods to get around this disappointingly generic feature of multi-field inflation were recently addressed in Ref.~\cite{JDD} and the special case of sum-separable potentials was also discussed in Ref.~\cite{Meyers}. In the case of sum-separable hilltop potentials which reach an adiabatic limit during inflation, what is known as the horizon-crossing approximation \cite{Nflation1} gives a good estimate of the final value for  $f_{\rm NL}$. 
\begin{equation}
f_{\rm{NL}}\approx -\frac{5}{6}\frac{V''_{\phi}}{V_{\phi}}\Bigg|_{*},
\end{equation}
where in this instance $\phi$ represents one or at most a few fields where $N_{i}$ is large. From this we see that provided there are enough fields to keep $\eta$ small, with the right initial conditions, a ridge can give rise to a large $f_{\rm NL}$. 

What we find in our analysis is that the problem of obtaining a large $f_{\rm NL}$ is made particularly acute by the need to obtain sufficient inflation. When we have a very smooth landscape, sufficient inflation is easily achieved but the lack of features means there is nothing to give rise to a large $f_{\rm NL}$. On the other hand, when the potential is very featured, it is difficult to start close enough to a ridge to get a large $f_{\rm NL}$ without falling off it, thereby killing inflation. In some models, such as axion N-flation \cite{Nfl}, a sufficiently large number of fields can make it possible to overcome this problem  \cite{Naxion}, as the large damping term makes it easier to be close to a ridge without falling off too soon. For our model, while we do not have the computational power to explore this possibility, with enough fields we would expect to see some examples with a large $f_{\rm NL}$, but they would constitute only a very small proportion. This is because if many fields have a large $N_{,i}$, their contributions to $f_{\rm NL}$ will, through a manifestation of the central limit theorem, cause $f_{\rm NL}$ to be vanishing in the limit of many contributing fields. Thus on average we would always expect $f_{\rm NL}$ to be small.

\begin{figure}[t]
\centering
\includegraphics[width=15cm]{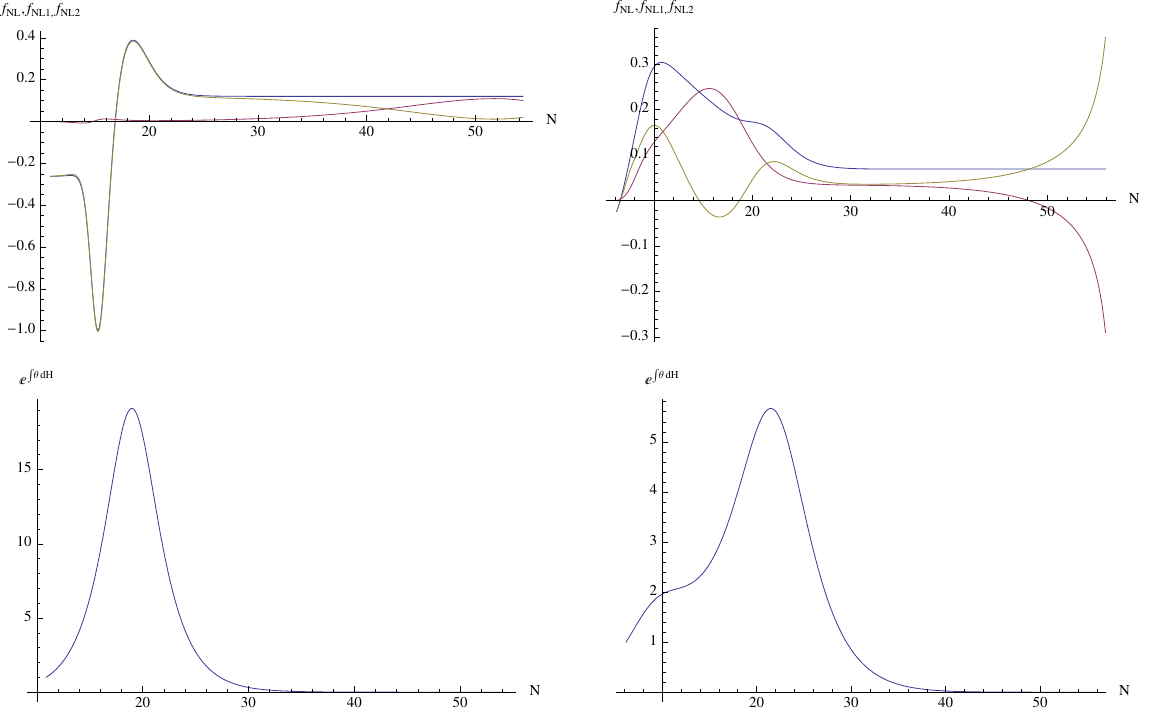}
\caption{Plots of Verse 113147 (left) and Verse 253911 (right) showing very distinctive  $f_{\rm NL}$ evolution. In the plots on the top row,  $f_{\rm NL}$ is blue,  the intrinsic component, $f_{\rm NL1}$, is red and the gauge contribution, $f_{\rm NL2}$, is yellow.}
\label{fig:examplefnls}
\end{figure}

\subsubsection{Trends in $f_{\rm NL}$ evolution}

While we found no examples of large $f_{\rm NL}$ we did find a very diverse range of behaviour. This diversity is indicative of why detection would be such a powerful constraint on models. That said we did find some common trends. Fig.~\ref{fig:examplefnls} shows two examples of evolution of $f_{\rm NL}$ together with the corresponding evolution in the width of the bundle. We chose these two examples, Verse 113147 and Verse 253911, in particular as each shows characteristics that were common to most trajectories, but also demonstrate that counter examples were found. 

\paragraph{The gauge contribution determines the peak}
As particularly well demonstrated by Verse 113147, in all examples we found $f_{\rm NL2}$ determined any peaks in $f_{\rm NL}$.  Typically it was the case that the intrinsic non-Gaussianity played a highly subdominant role in the feature, but we did find the exception that is Verse 253911 which clearly received an important contribution from the intrinsic part. In Ref.~\cite{MT1} it was noted that for the double quadratic potential and quadratic exponential potential this behaviour was present. Here we show this characteristic applies much more generally. 

\paragraph{$f_{\rm NL}$ grows when the bundle dilates}
Features in $f_{\rm NL}$ occur whenever $\Theta$ grows. This seems very reasonable since at this point the perturbed trajectories will be exploring different parts of the potential. Typically the peak in $f_{\rm NL}$ occurred very close to the time of the peak in $\Theta$ but again, Verse 253911 shows this need not be the case precisely. We see features in $f_{\rm NL2}$ are intimately related to features in $\theta$. We attribute this to common terms involving $V_{,ij}$.

\paragraph{Asymptotic behaviour of $f_{\rm NL}$ is not straightforward}
A result that continues to elude us is a simple way of  understanding what the final value of $f_{\rm NL}$ in the adiabatic limit will be. For sum-separable hilltop potentials, the horizon-crossing approximation works well, but for more general potentials there is no equivalent. As the examples in Fig.~\ref{fig:examplefnls} show, the asymptotic value can be reached in dramatically different manners. Worse still is the fact that in the adiabatic limit the intrinsic and gauge transformations need not settle to constant values. This indicates that a different set of parameters should be considered if we are to make progress with this question.

\subsection{Interlude: The Lyth bound}

Before moving on to discuss distributions of observables, we would like to take a brief moment to discuss the relation between field trajectories and the tensor-to-scalar ratio, as it will be helpful to bear in mind in the subsequent discussion. 

Taking $N_{\rm{CMB}}$ to be the number of e-folds between when fluctuations on CMB scales left the horizon and the end of inflation, we can obtain a $d$-field version of the Lyth bound by writing 
\begin{equation}
N_{\rm{CMB}}=\int_{\phi_{\rm{CMB}}}^{\phi_{\rm{end}}}\frac{1}{\sqrt{2\epsilon}}d\phi_{\parallel},
\end{equation}
where we are integrating along the field trajectory. If we assume $\epsilon$ is either constant or increasing over this period, then we have
\begin{equation}
2\epsilon< \frac{\Delta\phi}{N_{\rm{CMB}}}\,.
\end{equation}
For single field inflation $r=16\epsilon$ but when there are more fields the curvature perturbation evolves on super-horizon scales, suppressing $r$ and so $r<16\epsilon$ \cite{me}. We therefore see that the Lyth bound remains essentially the same for multi-field models as in the single-field case
\begin{equation}\label{eq:lythold}
r<16\epsilon<0.03\left(\frac{55}{N_{\rm{CMB}}}\right)^{2}\left(\frac{\Delta\phi}{M_{\rm{Pl}}}\right)^{2}
\end{equation}

Planck hopes to measure the tensor-to-scalar ratio with an accuracy of a few hundredths, hence has discovery potential if it is of order 0.1 or so. Comparing the Lyth bound with the discussion in section \ref{sec:fieldth} we therefore see that a detection would exclude all small- and medium-field models if only one field is admitted. However as previously discussed, multi-field models are strongly motivated by fundamental theory. If we consider the extreme case of sum-separable potentials then the discussion of Section \ref{sec:fieldth} requires each  $\Delta\phi_{i}\ll M_{\rm{Pl}}$ but there is no restriction on the number of fields contributing during inflation; this was for instance the motivation of the N-flation proposal \cite{Nfl}. Therefore, if we are to stay in the field theory favoured regime of small-field models, a detection of $r$ would place a lower bound on the number of fields! Rewriting Eq.~\eqref{eq:lythold} in a more suggestive form we have 
\begin{equation}
r<0.03d\left(\frac{55}{N_{\rm{CMB}}}\right)^{2} \frac{\Delta\phi_i\Delta\phi_i}{M_{\rm{Pl}}^2}\,.
\end{equation}

\subsection{Distribution of observables: $n$ and $r$}

\begin{figure}[t]
\centering
\includegraphics[width=15cm]{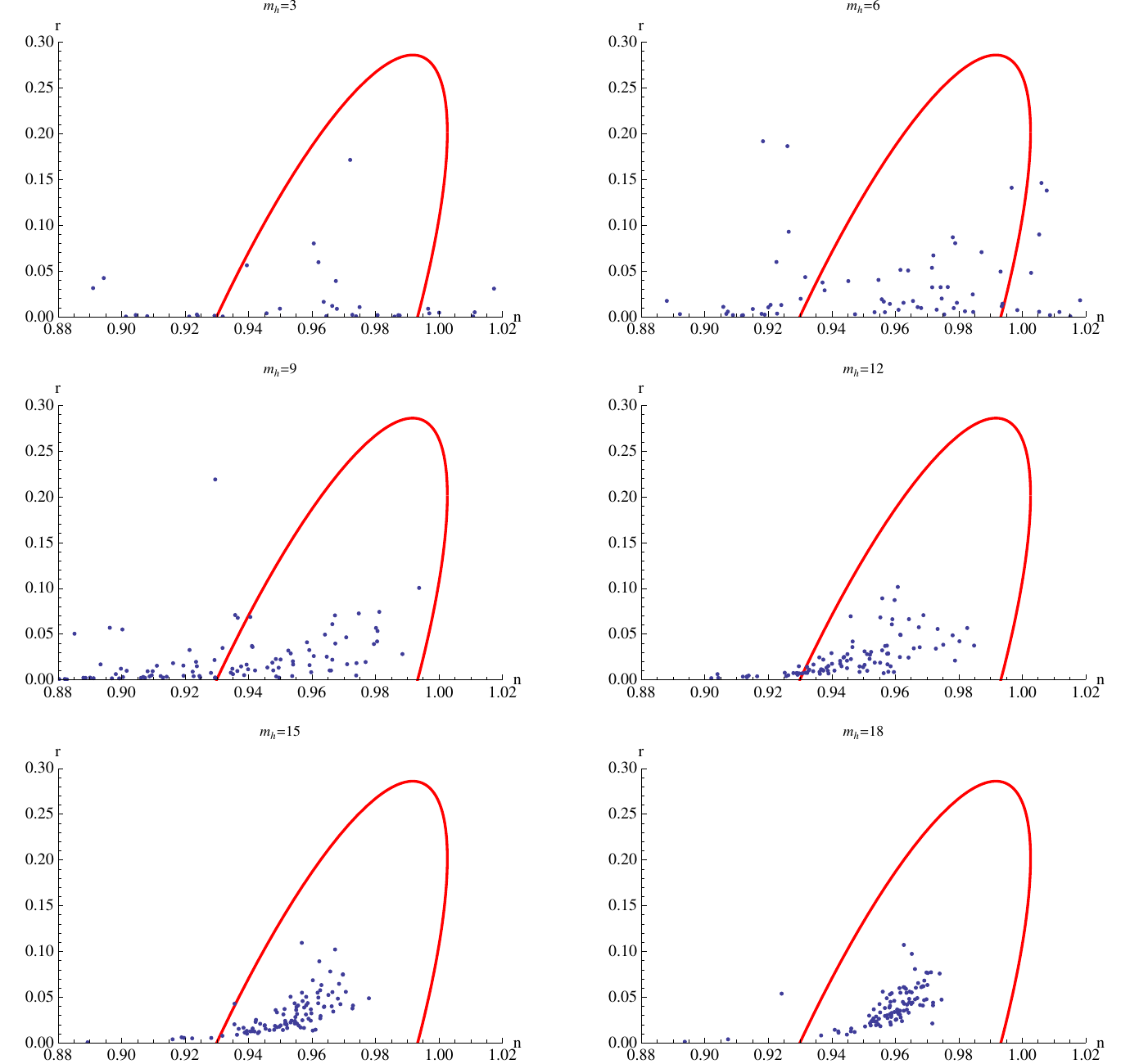}
\caption{$n$--$r$ plots for a range of $m_{{\rm h}}$, beginning with the more featured potentials. The curve shows the WMAP7+all 95\% confidence limit \cite{WMAP7}.}
\label{fig:nr}
\end{figure}

Fig.~\ref{fig:nr} shows our findings for $n$ and $r$ at the end of inflation, also summarised in Table \ref{tab:results}. As the landscape becomes more featured the viable trajectories become shorter and increasingly bendy. The Lyth bound tells us the distance travelled in field space places an upper bound on the tensor-to-scalar ratio. Furthermore, bends in the trajectory cause super-horizon evolution of the curvature power spectrum, while the tensor power spectrum is conserved, and so as Fig.~\ref{fig:nr} shows, we see an increasingly strong suppression in $r$ as we move to lower $m_{{\rm h}}$. 

\begin{figure}[t]
\centering
\includegraphics[width=14cm]{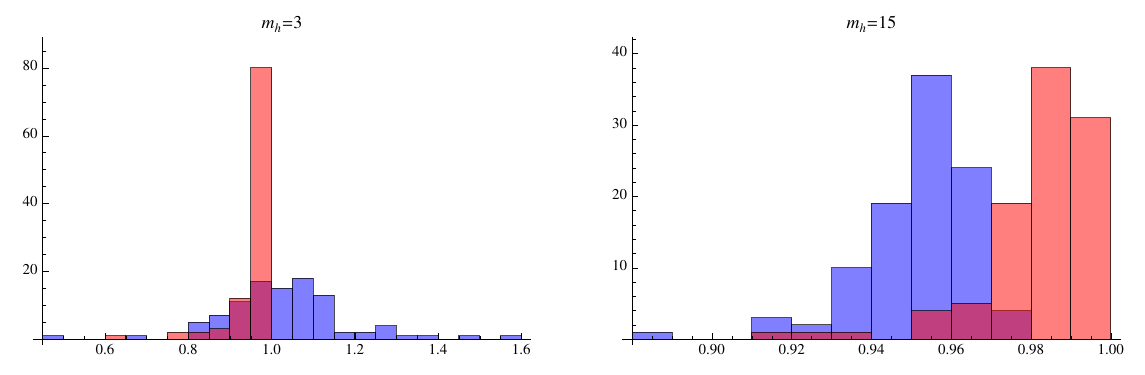}
\caption{Example plots comparing spectral index as obtained using single-field approximation (red) and as calculated taking evolution up to the end of inflation into account (blue).}
\label{fig:specexit}
\end{figure}

As summarised in Fig.~\ref{fig:Theta}, a less featured potential reduces the chance of trajectories reaching their adiabatic limit. It is noteworthy that, despite this, the plots of the $n$--$r$ plane for $m_{{\rm h}}=15\, M_{\rm{Pl}}$ and $m_{{\rm h}}=18\, M_{\rm{Pl}}$ show remarkable consistency for $n$ and $r$ at the end of inflation. This might lead one to think that the super-horizon evolution is having negligible effect, but if we take the example of $m_{{\rm h}}=15\, M_{\rm{Pl}}$, as Table \ref{tab:results} shows, the mean increase in the field trajectory from turning is only $4\%$, yet if we assumed a single-field approximation was valid we would obtain $n=0.98\pm0.01$ which is significantly different from the actual result of $0.95\pm0.01$. If nothing else, these results show one should be exceedingly careful when making single-field approximations. Fig.~\ref{fig:specexit} compares distributions for the spectral index with those obtained using a single-field approximation.

\begin{table}[t]
\centering
\begin{tabular}{|c|c|c|c|c|c|c|c|c|c|c|c|c|}
    \hline
  $m_{{\rm h}} (M_{\rm{Pl}})$ & $\Delta\phi$&$B_\phi (\%)$&$n$&$r$, $95\%$ conf.&$f_{\rm{NL}}$     \\ \hline
     3         &$4.3\pm1.7$ & $17\pm18 $&$1.03\pm 0.15$&$0.036$, $r<0.1981$&$-0.008\pm0.078$  \\ \hline
     6          &$7.6\pm1.8$& $16\pm18$&$0.93 \pm0.07$&$0.036$, $r< 0.172$&$0.029\pm0.034$ \\ \hline 
     9          &$ 8.4\pm1.8$& $10\pm13$&$0.93\pm0.03$&$0.023$, $r<0.081$ &$0.028\pm0.022$ \\ \hline
  12          &$8.6\pm1.4$& $5\pm7$&$0.94\pm0.02$&$0.024$, $r<0.066$&$0.022\pm0.010$    \\ \hline
  15          &$9.0\pm1.3$& $4\pm7$&$0.95\pm0.01$&$0.032$, $r<0.074$&$0.018\pm0.007$\\ \hline
   18         &$9.2\pm1.2$& $4\pm9$&$0.96\pm0.02$&$ 0.039$, $r< 0.079$&$0.016\pm0.009$  \\ \hline
    \end{tabular}
\caption{Table of the mean distance in field space travelled in the last 55 e-fold of inflation for a given $m_{{\rm h}}$, $B_\phi$, the mean percentage increase coming from bends in the trajectory, and corresponding results for observables.}
    \label{tab:results}
\end{table}

The central concern regarding the possible existence of a landscape is whether or not such a model can be tested. As we have mentioned, a key challenge is the measure problem, another being our very limited understanding of fundamental theory. Once these problems are better understood though, we will still be left with distributions for observable quantities. No matter how well developed our understanding, it seems reasonable to assume that at some level our ability to make predictions will be fundamentally limited by the details of the theory. Our toy landscape illustrates this in a very explicit way. In a sense $m_{{\rm h}}$ gives a way of quantifying the complexity of each landscape. For our model, we see the spread of results for the spectral index dramatically increases as we move to more featured landscapes, while for the tensor-to-scalar ratio the spreading is considerably less dramatic due to a suppression coming from the inevitable decrease in the length of the field trajectory $\Delta\phi$. As we have seen $f_{\rm NL}$ by contrast remains consistently small.
 
\section{Conclusions}
We explored inflationary dynamics in randomly-generated potentials as well as the consequences for super-horizon evolution of perturbations. We found this exploration to be interesting primarily on two fronts. 

First, by exploring a very large number of inflationary trajectories, we encountered a wide range of super-horizon evolution behaviour for $P_{\zeta\zeta}$ and $f_{\rm NL}$. The benefit of this was that it became easy to see what characteristics are generic and which are not. We found that peaks in $f_{\rm NL}$ tend to be determined by the gauge contribution but behaviour was rather more broad in the adiabatic limit, showing few trends. Understanding of non-Gaussianity is still in rapid development and so exploration of this kind can be very helpful in gaining insight in how to progress towards something more concrete such as Ref.~\cite{ray}.

We also found that keeping track of the easy-to-compute bundle width was extremely informative. By following how $\Theta$ changes along the trajectory, we were generally able to understand what qualities of the potential gave rise to super-horizon evolution. In particular, we found that peaks in $f_{\rm NL}$ occur during regions of the potential that give rise to a dilation of the bundle. However, a more quantitative description awaits future development. We emphasise that in order to make predictions in any multi-field model, one needs to perform an equivalent analysis to ensure no evolution is taking place at the time of evaluation. We found that as the mean length of the field trajectory in field space increased, the chances of reaching an adiabatic limit drastically decreased, rendering the larger field models essentially non-predictive without a model of reheating.

Second, we looked at how varying the scale of features and the number of light fields affected the ensembles produced for a given parameter. We found that landscapes where the mean length of the field trajectory was large typically gave results consistent with current observational data (despite not necessarily reaching their adiabatic limit). However for more featured landscapes where the mean field trajectory was smaller, the spread in the spectral index increased significantly.  The spread in the tensor-to-scalar ratio did not increase so dramatically. This can be understood in terms of the Lyth bound which places an upper bound on the tensor-to-scalar ratio according to the length of the field trajectory. We found varying the number of fields between 2 and 6 to have negligible effect on the distributions for observables.
Amongst all trajectories we found no examples of detectably large non-Gaussianity.  In absence of motivation for why we would be an atypical observer, this result is sufficiently strong to conclude that an observation of local-type non-Gaussianity by Planck would rule out models of this kind.

\acknowledgments
The authors were supported by the Science and Technology Facilities Council [grant numbers ST/1506029/1, ST/F002858/1, and ST/I000976/1]. We thank Mafalda Dias, Liam McAllister, David Mulryne, Sam Rogerson, and David Seery for numerous discussions relating to this work.

\end{document}